\documentclass[usenatbib,referee]{mn2e}
\usepackage{graphicx}

\usepackage{epsfig}
\usepackage[colorlinks,linkcolor=blue]{hyperref}
\usepackage{color}
\usepackage[dvipsnames]{xcolor}
\usepackage{amsmath}	
\usepackage{amssymb}	
\usepackage{subfigure}
\usepackage{soul}
\makeatletter

\newcommand{\Rmnum}[1]{\expandafter\@slowromancap\romannumeral #1@}
\makeatother
\usepackage{mathtools}
\usepackage{threeparttable}
\usepackage{booktabs}
\title[Formation of BWs]
{Formation of black widows through ultra-compact X-ray binaries with He star companions}

\author[Y. Guo, B. Wang and Z. Han]{
	Yunlang Guo,$^{\rm 1,2}$\thanks{E-mail:yunlang@ynao.ac.cn}
	Bo Wang$^{\rm 1,2}$\thanks{E-mail:wangbo@ynao.ac.cn} and
	Zhanwen Han$^{\rm 1,2}$\thanks{E-mail:zhanwenhan@ynao.ac.cn}
	\\
$^{1}$Yunnan Observatories, Chinese Academy of Sciences, Kunming 650216, China\\
$^{2}$University of Chinese Academy of Sciences, Beijing 100049, China\\
}

\date{Accepted XXX. Received YYY; in original form ZZZ}

\pubyear{2021}

\begin{document}
\label{firstpage}
\pagerange{\pageref{firstpage}--\pageref{lastpage}}
\maketitle

\begin{abstract}
Black widows (BWs) are a type of eclipsing millisecond pulsars (MSPs) with companion masses $M_2\lesssim 0.05\,\rm M_\odot$,
which can be used to study
the accretion history and the radiation of pulsars,
as well as the origin of isolated MSPs.
Recent observations indicate that there are two sub-types of BWs.
One is the BWs with $M_2$ $\sim$ $0.01-0.05\,\rm M_\odot$,
whereas another with $M_2$ $\lesssim 0.01\,\rm M_\odot$.
However,
the origin of the latter is still highly uncertain.
In this paper,
we investigated the formation of BWs with
$M_2$ $\lesssim 0.01\,\rm M_\odot$ through ultra-compact X-ray binaries (UCXBs) with He star companions,
in which a neutron star (NS) accretes material from a He star through Roche-lobe overflow.
By considering different He star masses and evaporation efficiencies with
the stellar evolution code Modules for Experiments in Stellar Astrophysics (MESA),
we evolved a series of NS+He star systems that can undergo UCXB stage.
This channel can explain the formation of the
BWs with $M_2$ $\lesssim 0.01\,\rm M_\odot$
within the Hubble time,
especially three widely studied BWs, i.e. PSRs J1719-1438, J2322-2650 and J1311-3430.
We found that X-ray irradiation feedback does not affect the evolutionary tracks of evaporation process.
The simulations indicate that the UCXB channel with He star companions are the potential progenitors of isolated MSPs,
and that the origin of BWs with $M_2$ $\lesssim 0.01\,\rm M_\odot$
is different with another sub-type of BWs.
In addition,
the present work suggests that the BWs with $M_2$ $\lesssim 0.01\,\rm M_\odot$
may not be produced by redback systems.

\end{abstract}

\begin{keywords}
binaries: close -- stars: evolution -- pulsars: general.
\end{keywords}

\section{Introduction}
It is generally believed that millisecond pulsars (MSPs) are
old radio pulsars with spin periods $\textless 30$\,ms
(e.g. Backer et al. 1982; Lorimer 2008).
They are proposed to be formed from the evolution of intermediate-mass X-ray binaries
and low-mass X-ray binaries
(LMXBs; e.g. Tauris \& van den Heuvel 2006; D'Antona \& Tailo 2020).
The neutron stars (NSs) can be spun up to MSPs
by accreting the material and angular momentum from their companions,
called the recycling scenario
(e.g. Alpar et al. 1982; Bhattacharya \& van den Heuvel 1991; Tauris, Langer \& Kramer 2012).
The surface magnetic field of NSs will decay from $10^{12}$\,G to $10^8-10^9$\,G during the recycling process,
although the decay mechanism is still unclear
(e.g. Srinivasan et al. 1990; Konar \& Bhattacharya 1997; Cumming, Zweibel \& Bildsten 2001).
The eclipsing MSPs have orbital periods ranging from $\sim 0.1$ to $1.0$ days,
which can be mainly divided into two types,
i.e. black widows (BWs) and redbacks
(RBs; see Roberts 2013).
BWs have companions with masses $\lesssim 0.05\,\rm M_\odot$,
while RBs have companions with masses in the range of $0.1-1.0\,\rm M_\odot$
(e.g. Bailes et al. 2011; Nieder et al. 2020).

MSPs are important objects for studying
the evolutionary history of NS systems,
such as the accretion efficiency, the accretion disk structure,
the magnetic braking, and the decay of the surface magnetic field of NSs, etc
(e.g. Srinivasan et al. 1990; van Paradijs 1996; Tauris 2011; Antoniadis et al. 2012;
Van, Ivanova \& Heinke 2019; Chen et al. 2021; Wang, Liu \& Chen 2022).
Especially, the maximum accreted mass of NSs can be determined by the accretion physics,
which can be used to provide constraints on their mass distribution and equation of state
(e.g. Kiziltan et al. 2013; Alsing, Silva \& Berti 2018; Godzieba, Radice \& Bernuzzi 2021; Li et al. 2021).
In addition, about one-third of MSPs are isolated in the observations,
but their formation mechanism is still unclear (e.g. van den Heuvel \& Bonsema 1984; Verbunt et al. 1987;
Freire et al. 2011; Jiang et al. 2020).
It has been suggested that
BWs may be the evolutionary link between the accreting X-ray pulsars and the isolated MSPs,
in which the companion is ablated by the $\gamma$-ray and the energetic particles emitted by MSPs
(e.g. Kluzniak et al. 1988; van den Heuvel \& van Paradijs 1988).
This idea is supported by the discovery of an original BW,
i.e. B1957+20, an eclipsing 1.6 ms pulsar
in a 9.17\,h circular orbit with a low mass companion ($\sim0.02\rm\,M_\odot$; Fruchter, Stinebring \& Taylor 1988).
In the standard LMXB channel,
a NS accretes
material from a main-sequence (MS) star through Roche lobe
overflow
(RLOF; e.g. Tauris \& Savonije 1999; Podsiadlowski, Rappaport \& Pfahl 2002; Chen, Liu \& Wang 2020).
However,
the formation of isolated MSPs cannot be explained
by this channel within the Hubble time 
(e.g. Chen et al. 2013; Ginzburg \& Quataert 2020).

In order to reproduce the eclipsing MSPs,
the following conditions need to be considered.
(1) The NS spins with a millisecond period.
(2) The pulsar becomes a radio MSP because of the low mass-transfer rate.
(3) The companion is ablated by pulsar radiation.
(4) The orbital period and the companion mass ($M_2$) in the simulations are consistent with the observations.
Up to now, several channels for the formation of BWs have been proposed, as follows:
(1) The pulsar begins to evaporate the companion when the magnetic braking ceases
and the companion detaches from its Roche lobe
(e.g. Chen et al. 2013; Jia \& Li 2015).
(2) The evaporation process occurs when the mass transfer ceases
due to the cyclic mass transfer caused by the X-ray irradiation feedback
(e.g. B{\"u}ning \& Ritter 2004; Benvenuto, De Vito \& Horvath 2012, 2014).
(3) A MSP is formed once an oxygen-neon white dwarf (WD) undergoes accretion-induced collapse,
and then begins to evaporate its companion 
(e.g. Smedley et al. 2015; Liu \& Li 2017; Liu et al. 2018; Ablimit 2019; Wang 2018; Wang \& Liu 2020).

Recent observations indicate that
BWs can be divided into two sub-types:
one is the BWs with $M_2$ $\sim0.01-0.05\,\rm M_\odot$,
while another sub-type of BWs have $M_2$ $\lesssim 0.01\,\rm M_\odot$
(e.g. Keith et al. 2010; Ray et al. 2012).
The BWs with $M_2$ $\sim0.01-0.05\,\rm M_\odot$ can be explained by the pulsar radiation evaporating MS stars
(e.g. Chen et al. 2013; Benvenuto, De Vito \& Horvath 2014; Jia \& Li 2015).
However,
the BWs with $M_2$ $\lesssim 0.01\,\rm M_\odot$ are hardly
reproduced by previous studies (e.g. Chen et al. 2013; Ginzburg \& Quataert 2021),
especially three widely studied BWs,
i.e. PSRs J1719-1438, J2322-2650 and J1311-3430, as follows:
(1) PSR J1719-1438 has a companion with mass close to Jupiter
($\sim 1.16\times10^{-3}\rm M_\odot$),
and the composition of its companion may be helium or carbon-oxygen (see Bailes et al. 2011).
It has been suggested that PSR J1719-1438 may be formed from ultracompact X-ray binaries
(UCXBs; e.g. Nelemans \& Jonker 2010; Nelemans et al. 2010; Heinke et al. 2013; Wang et al. 2021),
but the standard UCXB channel with MS star companions hardly reproduces this system within the Hubble time
(e.g. Bailes et al. 2011; van Haaften et al. 2012).
(2) Similar to PSR J1719-1438, PSR J2322-2650 has a planetary-mass companion star ($\sim 10^{-3}\rm M_\odot$),
and its orbital period is 7.75\,h (see Spiewak et al. 2018).
Such a low-mass companion implies that its formation mechanism may be similar to that of PSR J1719-1438.
(3) PSR J1311-3430 is a BW with a low-mass companion ($\sim 10^{-2}\rm M_\odot$).
This BW has an orbital period of 93 min and the minimum mean density of the companion is $45\rm\,g\,cm^{-3}$,
which means that this system may originate from the UCXBs (e.g. Pletsch et al. 2012; Romani et al. 2012).
Additionally, the companion of PSR J1311-3430 has a helium-dominated photosphere,
indicating that its companion may be a He star (e.g. Romani et al. 2012; Romani, Filippenko \& Cenko 2015).

Wang et al. (2021) studied the formation and evolution of UCXBs with He star companions systematically.
They found that this channel can explain the formation of five transient sources with relatively long orbital periods.
However, they did not consider the evaporation process that may occur during the evolution of NS+He star systems.
When the mass-transfer rate in NS+He star system drops to less than a critical value on the declining stage
(see Fig.\,4 in Wang et al. 2021),
this system will appear to be a transient source due to the thermal-viscous instability of accretion disks
(e.g. King, Kolb \& BurderiLasota 1996; Dubus et al. 1999; Lasota et al. 2008).
At this moment,
the NS may turn on pulsar radiation and
begin to evaporate its companion.

Following the work of Wang et al. (2021),
the purpose of this paper is to explore the
origin of BWs with $M_2$ $\lesssim 0.01\,\rm M_\odot$
by considering the evaporation process based on the UCXB channel.
In Section 2, we described the numerical methods and assumptions for the binary evolution simulations.
The results are presented in Section 3.
The discussions are given in Section 4.
Finally, we summarize the results in Section 5.

\section{Numerical methods and assumptions}

We carried out detailed binary evolution calculations of NS+He star systems that can undergo UCXB stage by using
the stellar evolution code Modules for Experiments in Stellar Astrophysics
(MESA, version 12778; see Paxton et al. 2011, 2013, 2015, 2018, 2019).
In our simulations,
the basic physical assumptions are similar to those of Wang et al. (2021).
We performed the long-term evolution of
four NS+He star systems
located in the UCXB parameter space (see Fig.\,8 in Wang et al. 2021),
in which we set the initial NS mass ($M_{\rm NS}^{\rm i}$) to be $1.4\,\rm M_\odot$
and adopted different initial orbital periods,
i.e. log$P_{\rm orb}^{\rm i}$\,(d) = $-1.90, -1.80, -1.65$ and $-1.55$.
The initial masses of He star companions ($M_{2}^{\rm i}$) are in the range of $0.32-0.60\,\rm M_\odot$,
in which $0.32\,\rm M_\odot$ is the minimum mass required for a He star to ignite the central helium
(see Han et al. 2002).

During the binary evolution,
the He star fills its Roche lobe due to the orbital contraction induced by the radiation of gravitational wave (GW).
In the present work,
we assume that the NSs are point masses,
and the composition of He star is $98\%$ helium and $2\%$ metallicity.
Following the method described in Podsiadlowski, Rappaport \& Pfahl (2002),
we set the fraction of transferred material accreted by NS to be $0.5$.
We adopted an Eddington accretion
rate, $\dot M_{\rm Edd}=3 \times 10^{-8}\,\rm M_\odot \rm yr^{-1}$,
which can be used to limit the mass-accretion rate (e.g. Dewi et al. 2002; Chen, Li \& Xu 2011).
Similar to previous studies,
we neglected the effect of magnetic braking on the evolution of NS+He star systems
(e.g. Iben \& Tutukov 1987; Iben et al. 1987; Yoon \& Langer 2003; Tang, Liu \& Wang 2019; Wang et al. 2009,2021).
This is because the magnetic braking is usually used for Sun-like stars that have a convective envelope and a radiative core
(e.g. Rappaport, Verbunt \& Joss 1983; Paxton et al. 2015; Deng et al. 2021).
In addition, we did not consider the angular momentum loss caused by evaporation wind.

We suppose that the He star companion begins to be evaporated by the pulsar radiation
when the mass-transfer rate is lower than $\sim 5 \times 10^{-10}\,\rm M_\odot \rm yr^{-1}$
on the declining stage.\footnote{Lasota
	et al. (2008) pointed out that for irradiated pure He disks,
	the critical mass-transfer rate leading to the thermal-viscous instability of accretion disks
is about $5 \times 10^{-10}\,\rm M_\odot \rm yr^{-1}$ at the orbital period around $25$ min.}
The mass-loss rate of companion star driven by pulsar radiation is expressed as follows
(see Stevens, Rees \& Podsiadlowski 1992):
\begin{equation}
\dot M_{\rm 2, evap} = -\frac{f}{2v_{\rm 2, esc}^2}L_{\rm P}(\frac{R_2}{a})^2,
\end{equation}
where $v_{\rm 2, esc}$ is the escape velocity at the surface of the
companion star,
$R_2$ is the radius of the companion star,
and $a$ is the the orbital separation of the binary.
$f$ is the evaporation efficiency,
which is an undetermined factor (e.g. Ruderman et al. 1989a,b; Abdo et al. 2013; Ginzburg \& Quataert 2020).
In this work, we assume that the companion is evaporated by the pulsar radiation with $f\le0.1$
(e.g. Chen et al. 2013; Jia \& Li 2015).
$L_{\rm P}$ is the spin-down luminosity, which can be given by $L_{\rm P}=4\pi^2I\dot P_{\rm spin}/P_{\rm spin}^3$,
where $I$ is the pulsar moment of inertia,
$P_{\rm spin}$ and $\dot P_{\rm spin}$ are the spin period and the spin-down rate of NSs, respectively.
In our calculations,
we set $I = 10^{45}\,\rm g\,\rm cm^2$, the initial spin period for MSPs $P_{\rm spin}^{\rm i}=3\rm\,ms$,
the initial period derivative $\dot P_{\rm spin}^{\rm i}=1.0\times10^{-20}\rm\,s\,s^{-1}$,
and the constant braking index $n=3$ in the standard magnetic dipole radiation model
(e.g. Chen et al. 2013; Jia \& Li 2015).

\section{Results}
We performed a series of calculations with different initial He star masses
and initial orbital periods.
Table 1 lists the evolutionary properties of NS+He star binaries before the evaporation process.
\begin{table*}
	\centering
	
	\caption{The evolutionary properties of NS+He star systems with different initial companion masses and initial orbital periods before the evaporation process.
		$M_2^{\rm i}$ and log$P_{\rm orb}^{\rm i}$ are the initial He star mass in solar masses and the initial orbital period in days;
		$t_{\rm evap}$, $M_{\rm NS, evap}$, $M_{\rm 2, evap}$ and $P_{\rm orb, evap}$ are the age of binary system, the NS mass,
		the companion mass and the orbital period at the beginning of evaporation process, respectively.}
	\begin{tabular}{ l  c c c ccc c c c  l }
		\toprule
		\hline 
		Set	&$M_2^{\rm i}$  &log$P_{\rm orb}^{\rm i}$  &$t_{\rm evap}$ & $M_{\rm NS, evap}$ &$M_{\rm 2, evap}$& $P_{\rm orb, evap}$\\
		&($\rm M_\odot$)   	&(d)	  					& (Myr)       &	($\rm M_\odot$) &$(\rm M_\odot)$& (min)\\
		\hline 
		1	&$0.32$			&$-1.90$			&$37.41$		&$1.49$ &$0.058$&$28.26$\\
		2	&$0.40$ 		&$-1.80$			&$38.70$		&$1.56$ &$0.053$&$27.08$\\
		3	&$0.50$ 		&$-1.65$			&$42.60$		&$1.60$ &$0.055$&$28.75$\\
		4	&$0.60$ 		&$-1.55$			&$47.66$		&$1.66$ &$0.054$&$28.37$\\
		\hline 
	\end{tabular}
\end{table*}
\subsection{Evolutionary tracks}
\begin{figure*}
	\centering\includegraphics[width=\columnwidth*4/5]{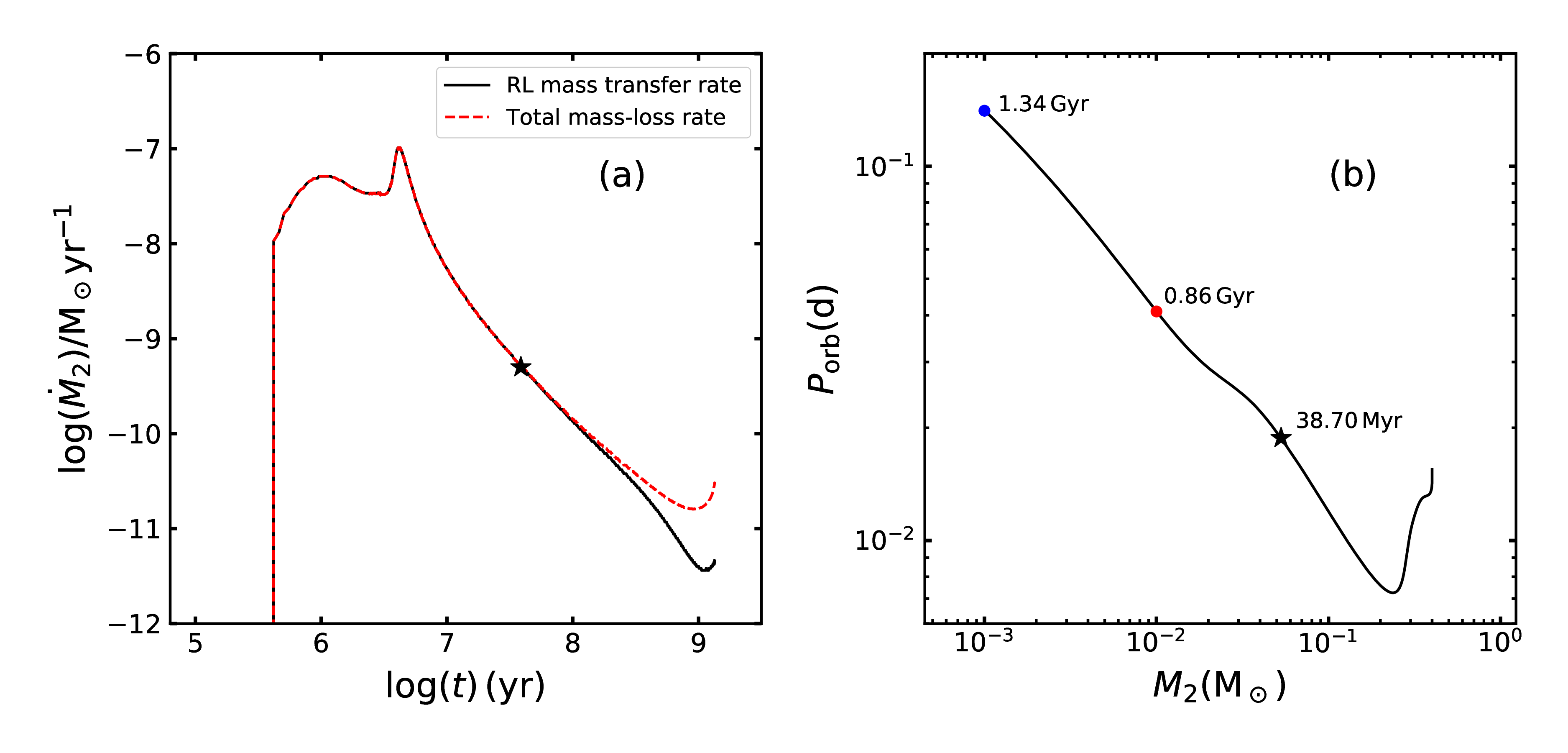}
	\caption{A representative example of the evolutionary track of UCXBs with He star companions,
		in which $M_{\rm NS}^{\rm i}=1.4\,\rm M_\odot$, $M_2^{\rm i}=0.4\,\rm M_\odot$,
		log$P_{\rm orb}^{\rm i}(\rm d)=-1.80$ and $f=0.01$.
		Panel (a): Evolutionary track of mass-transfer rate (black line) and total mass-loss rate (red, dashed line).
		Panel (b): Orbital period vs. donor mass during the binary evolution.
		The beginning of evaporation process is marked with the black stars.
		The red and blue dots denote the moments when the donor mass decreases to $0.01\,\rm M_\odot$ and $0.001\,\rm M_\odot$, respectively.}
	\label{fig:he0.01}
\end{figure*}

\begin{figure*}
	\centering\includegraphics[width=\columnwidth*4/5]{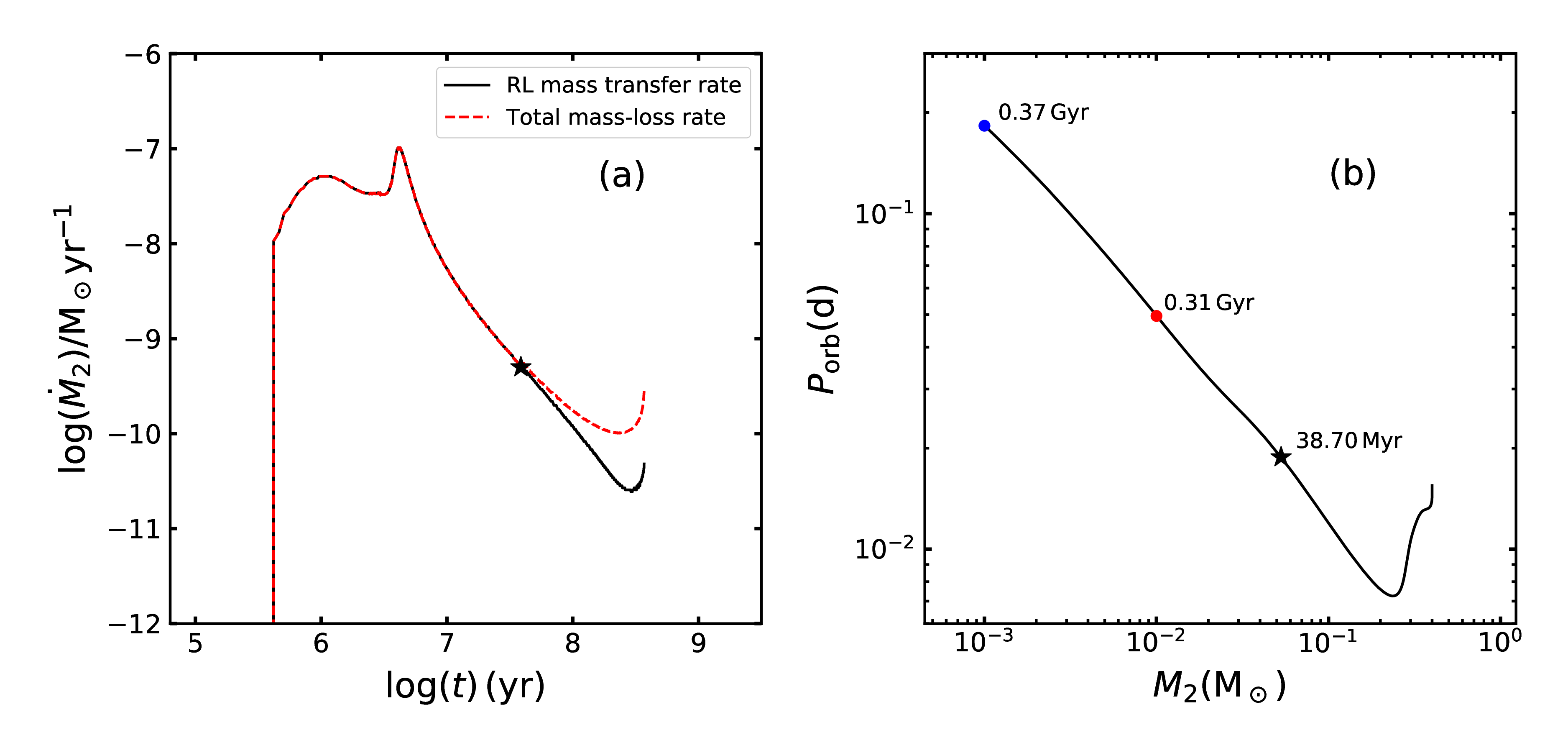}
	\caption{Similar to Fig.\,1, but for $f = 0.07$.}
	\label{fig:he0.07}
\end{figure*}

\begin{figure*}
	\centering\includegraphics[width=\columnwidth*4/5]{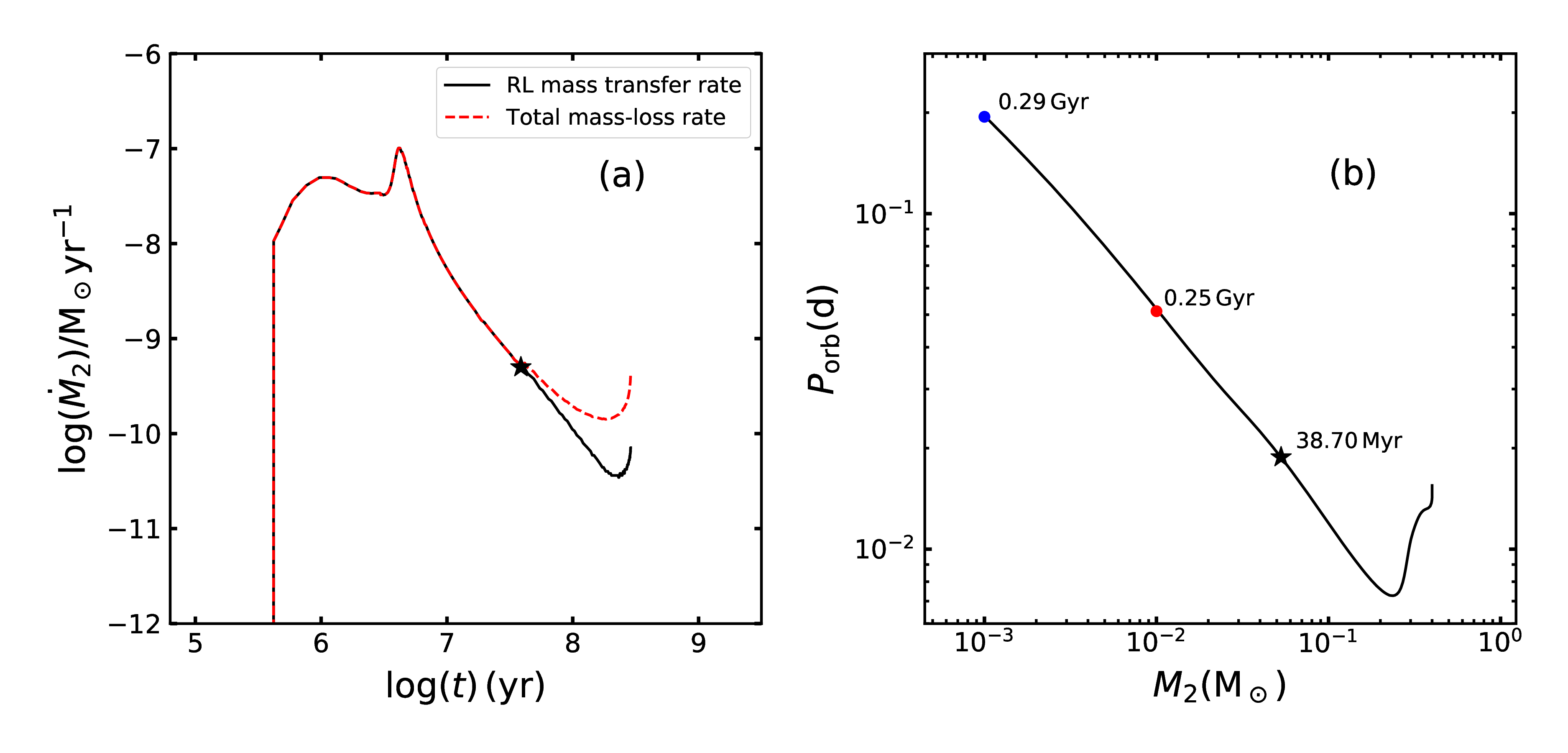}
	\caption{Similar to Fig.\,1, but for $f = 0.1$.}
	\label{fig:he0.1}
\end{figure*}
Figs\,$1-3$ show a representative evolutionary example of the NS+He star system
with $M_2^{\rm i}=0.40\,\rm M_\odot$, log$P_{\rm orb}^{\rm i}$(d)$=-1.80$.
Similar to previous studies,
we adopted three typical evaporation efficiencies, i.e. $f=0.01, 0.07$ and $0.1$
(see, e.g. Chen et al. 2013).
The NS+He star systems undergo two stages
(i.e. the pre-evaporation stage and the evaporation stage),
which are divided by the black stars in these figures.

During the pre-evaporation stage,
the binaries evolve into persistent UCXBs
once the NSs accrete material from the He star companions through the inner Lagrangian point (see Wang et al. 2021).
The binary reaches the minimum orbital period
as the rapid shrinking of the orbital separation due to the angular momentum loss induced by the GW radiation.
At this stage, the companion star decreases its mass to $\sim0.24\,\rm M_\odot$,
and the helium burning at the center of the He star begins to fade gradually.
It is worth noting that
the He stars gradually reach a mildly degenerate state
when the binaries experience the minimum orbital period
(see Fig.\,6 in Wang et al. 2021).
After about 38.70\,Myr,
the mass-transfer rate decreases to $\sim 5 \times 10^{-10}\,\rm M_\odot \rm yr^{-1}$,
resulting in the thermal-viscous instability of accretion disks.
At this moment,
the NS increases its mass to $1.56\,\rm M_\odot$
and spins up to become a MSP,
causing the companion star to be evaporated by the pulsar radiation.
Meanwhile,
the orbital period increases to $27.08$\,min,
and the companion mass decreases to $0.053\,\rm M_\odot$ (see Table 1).

During the evaporation stage,
the binary orbits gradually widen,
and the evaporation wind would increase the mass-loss rate of companion
(see the red dashed line in Figs\,$1-3$).
The accretion process stops once the pulsar radiation turns on,
since the radiation pressure would drive away the transferred material near the inner Lagrangian point
(e.g. Kluzniak et al. 1988; Burderi et al. 2001).
The $f$-value will affect the evolutionary track of the binary system when the evaporation process begins.
We note that
the He star mass will decrease faster and the binary orbits widen more quickly
as the $f$-value increases,
owing to the more efficient evaporation.
In addition,
Figs\,$1-3$ represent the binary ages
when the mass of He star decreases to $0.01\,\rm M_\odot$ and $0.001\,\rm M_\odot$, respectively.
The binary age would not exceed the Hubble time
when the companion mass reaches the planetary-mass ($\sim0.001\,\rm M_\odot$).

\begin{figure*}
	\centering\includegraphics[width=\columnwidth*5/5]{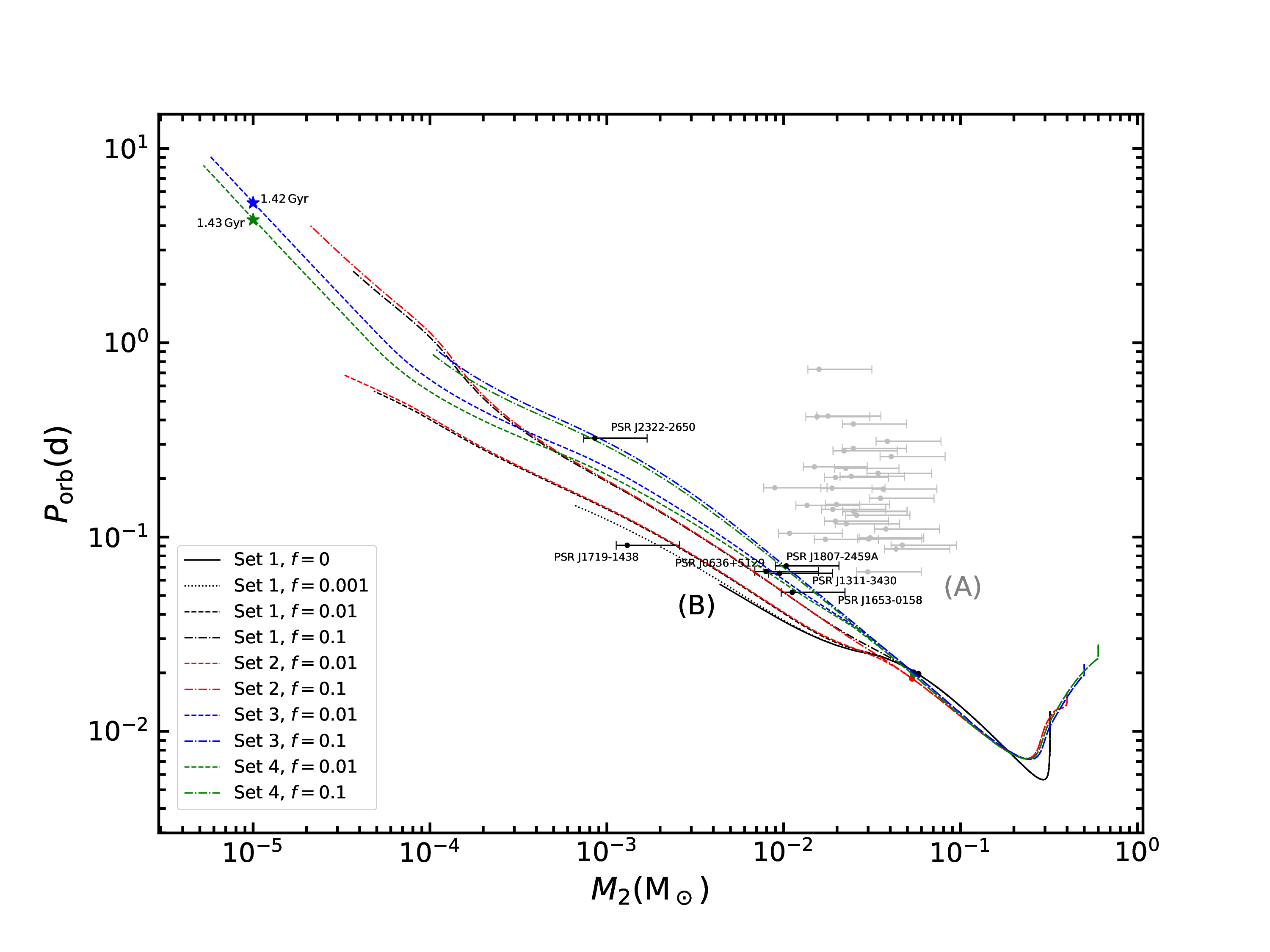}
	\caption{The evolutionary tracks of NS+He star systems with different
		initial He star masses and $f$-values,
		in which these systems can undergo the UCXB stage.
		Different colored lines represent sets $1-4$ in Table 1, and different types of lines represent different $f$-values.
		Different colored dots indicate the moment when the evaporation process starts in different sets.
        The stars denote the moment when the companion mass decreases to $10^{-5}\,\rm M_\odot$.
	The observed data of BWs are taken from the ATNF
Pulsar Catalogue \href{http://www.atnf.csiro.au/research/pulsar/psrcat}{http://www.atnf.csiro.au/research/pulsar/psrcat}
(version 1.67, December 2021; Manchester et al. 2005),
and P. Freire's Web site \href{http://www.naic.edu/$\sim$pfreire/GCpsr.html}{http://www.naic.edu/$\sim$pfreire/GCpsr.html}.
The minimum and maximum of error bars correspond to an orbital inclination angle of $90^{\circ}$ and $26^{\circ}$ (the $90\%$
probability limit), respectively.
The black samples in region (B) represent the BWs that this work can explain,
and the gray samples in region (A) represent other BW systems.
}
	\label{fig:pmz4}
\end{figure*}

\subsection{Comparison with observations}
To compare with observations,
we simulated a number of complete binary computations for the NS+He star systems
listed in Table 1 by considering different $f$-values,
in which we adopted four $f$-values in set $1$
to study the effect of $f$-value on the evolutionary tracks,
i.e. $f = 0, 0.001, 0.01$ and $0.1$.
Fig.\,4 represents the evolutionary tracks of ten representative binaries
in the orbital period versus the donor mass diagram,
as well as BW samples in the observations.
The BW samples can be divided into two sub-types:
(1) The BWs in region (A) have companions with masses $M_2$ $\sim0.01-0.05\,\rm M_\odot$,
which can be reproduced by the pulsar radiation evaporating MS stars
(e.g. Chen et al. 2013; De Vito \& Horvath 2014; Jia \& Li 2015).
(2) The BWs in region (B) have companions with masses $\lesssim0.01\,\rm M_\odot$,
including the BWs with $M_2\sim0.001\,\rm M_\odot$
and the relatively compact BWs with $M_2\sim0.01\,\rm M_\odot$.

As shown in Fig.\,4,
the binary evolutionary tracks pass through the region (B) of BWs with $M_2$ $\lesssim 0.01\,\rm M_\odot$,
i.e. PSRs J1719-1438, J2322-2650, J1311-3430, J1807-2459A, J0636+5129 and J1653-0158.
This indicates that 
the BWs with $M_2$ $\lesssim 0.01\,\rm M_\odot$ can be explained
by the UCXB channel with He star companions.
Meanwhile, we note that the relatively short period BW systems with planetary-mass companion
are preferentially formed from the binaries with less massive He stars, e.g. PSR J1719-1438.
In addition,
the evolutionary track represented by the black solid line (i.e. set 1, no evaporation)
can be used as the boundary for this channel to form BWs,
since $0.32\,\rm M_\odot$ is the minimum mass of He stars that can ignite the central He.
It is worth noting that
the BW companions are helium-rich and mildly degenerate.
\begin{table*}
	\centering
	\caption{Models versus observations of five BW samples.
			The columns (from left to right): the initial companion mass, the initial orbital period, the evaporation
			efficiency, and the initial spin period of pulsars; the constant $K$ of pulsars,
			the companion mass and the orbital period of BWs; the spin period, the spin-down rate, the spin-down luminosity of pulsars, and the binary age.
			We did not list PSR J1807-2459A in this table, because it has a negative $\dot P_{\rm spin}$ of $-4.34\times10^{-21}$\,s\,$\rm s^{-1}$.
	}
	\begin{tabular}{ lccccccccccccl }
		\toprule
		\hline \\
		Pulsars/Model&& $M_2^{\rm i}$&log$P_{\rm orb}^{\rm i}$&$f$&$P_{\rm spin}^{\rm i}$&$K$& $M_2$ &$P_{\rm orb}$ &$P_{\rm spin}$ &$\dot P_{\rm spin}$ &$L_{\rm P}$&$t$ \\
		&&    ($\rm M_\odot$)& (d) & &(ms)&(s)& ($10^{-2}\,\rm M_\odot$)& (d) &(ms)&(s\,$\rm s^{-1}$)&($L_\odot$)&(Gyr)\\
		\hline \\
		J1719-1438&& ...&...&... &... &4.7e-23&$0.13_{-0.02}^{+0.13}$ & 0.090& 5.8& 8.04e-21& 0.41 & ...\\
		Model\,1	  &&  0.32 & $-1.90$& 0.002&3.3&4.7e-23& 0.19& 0.090& 5.7& 8.14e-21&0.45& 7.48\\
		\hline \\
		J2322-2650&& ...&...&... &... &2.0e-24&$0.086_{-0.01}^{+0.08}$ & 0.323& 3.50& 5.83e-22& 0.14 & ...\\
		Model\,2    &&0.50 & $-1.65$& 0.7&3.3&2.0e-24& 0.076& 0.323& 3.31& 6.10e-22& 0.17& 0.59 \\
		\hline \\
		J1311-3430&& ...&...&...  &... &5.4e-23&$0.95_{-0.13}^{+0.93}$ & 0.065& 2.6& 2.10e-20& 12.83 & ...\\
		Model\,3  &&0.50 & $-1.65$& 0.003&2.0&5.4e-23& 0.94& 0.065&2.6 &2.08e-20 & 12.35& 0.84 \\
		\hline \\
		J0636+5129&& ...&...&...  &... &9.9e-24&$0.80_{-0.11}^{+0.78}$ &0.067& 2.9& 3.45e-21& 1.50 & ...\\
		Model\,4   && 0.60 & $-1.55$& 0.03&2.8&9.9e-24&0.83& 0.067& 2.9& 3.41e-21& 1.44& 0.95 \\
		\hline \\
		J1653-0158&& ...&...&...  &... &4.7e-23&$1.12_{-0.15}^{+1.10}$ &0.052& 2.0& 2.40e-21& 3.24 & ...\\
		Model\,5 &&0.60 & $-1.55$& 0.01&1.9&4.7e-23& 1.18& 0.052& 2.0& 2.40e-21& 3.24& 0.93 \\
		\hline
	\end{tabular}
\end{table*}

To further reproduce the properties of BWs
(e.g. $P_{\rm spin}$, $\dot P_{\rm spin}$ and $L_{\rm P}$),
we considered different input parameters for initial models by using the method described in Chen (2017).
We assume that pulsars are spinning down based on a power law $\dot\nu = -K\nu^3$
due to a pure magnetic dipole radiation,
in which $\nu$ and $\dot\nu$ are the spin frequency and its derivative, respectively.
$K$ is a constant depending on the momentum of inertia, the magnetic field, and the radius of the pulsar.
We derived the constant $K$ for each source based on the measured $\nu$ and $\dot \nu$.
Besides, we also adopted different $P_{\rm spin}^{\rm i}$.

Table 2 shows the comparison for the relevant properties of BWs
between our simulations and the observed samples,
including $M_2$, $P_{\rm orb}$, $P_{\rm spin}$, $\dot P_{\rm spin}$ and $L_{\rm P}$.
The BW properties can be well reproduced by using appropriate $f$-values and $P_{\rm spin}^{\rm i}$.
It is worth noting that a high $f$-value is needed to explain PSR J2322-2650
due to the low $L_{\rm P}$ for this source.
We speculate that PSR J2322-2650 may also be produced by
the UCXB channel with a more massive companion and a low $f$-value.
However, we suffered from some numerical difficulties
when the companion mass is $\textgreater 0.6\,\rm M_\odot$ (see also Wang et al. 2021).
In addition, PSRs J1719-1438 and J2322-2650 have the lowest $L_{\rm P}$ among the observed BWs.
A low value of $L_{\rm P}$ will reduce the mass-loss rate of the companion star caused by the pulsar radiation,
slowing down the evolution of PSRs J1719-1438 and J2322-2650 towards isolated MSPs.
\section{Discussion}
\subsection{X-ray irradiation feedback}
During the RLOF phase in LMXBs,
X-ray will illuminate the companion
when the material falls on the surface of the NS,
and then affects the evolution of binary systems
(e.g. Podsiadlowski 1991; B{\"u}ning \& Ritter 2004; Benvenuto, De Vito \& Horvath 2012, 2014).
In order to explore the effect of the X-ray radiation heating companion on the final results,
we further considered the X-ray irradiation feedback based on the UCXB channel.
\begin{figure*}
	\centering\includegraphics[width=\columnwidth*4/5]{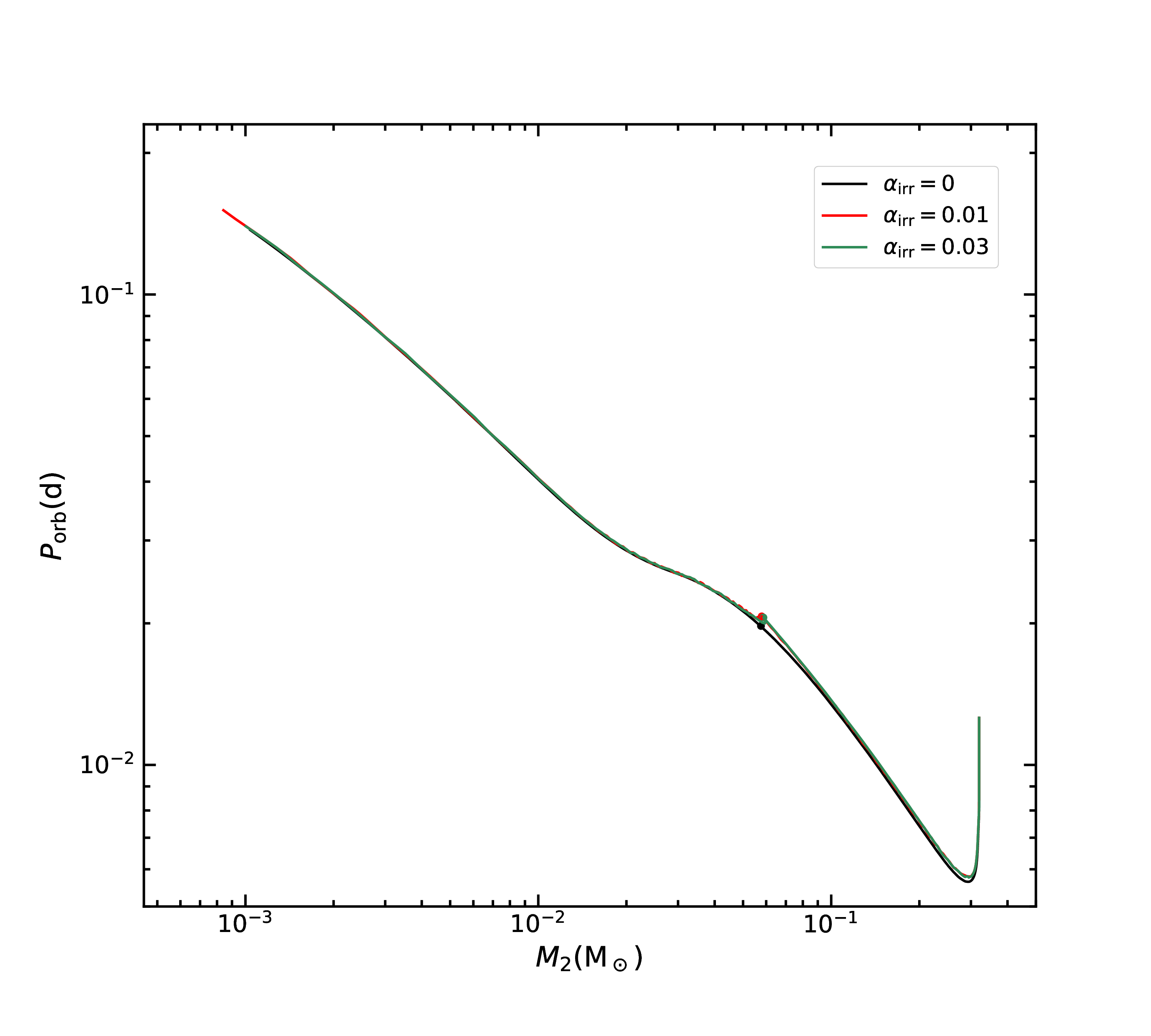}
	\caption{The evolutionary track of NS+He star systems with different irradiation efficiencies,
		in which the evaporation efficiency is $0.01$ and the initial He star mass is $0.32\,\rm M_\odot$.
		These dots denote the moment when the evaporation turns on and the X-ray irradiation feedback ends.}
	\label{fig:X-raybwx}
\end{figure*}

The X-ray irradiation flux ($F_{\rm irr}$) can be calculated as follows:
\begin{equation}
F_{\rm irr} = \alpha_{\rm irr}\frac{L_{\rm acc}}{4\pi a^2},
\end{equation}
where $\alpha_{\rm irr}$ is the irradiation efficiency ranging from $0.01$ to $0.1$ in LMXBs
(e.g. B{\"u}ning \& Ritter 2004; Benvenuto, De Vito \& Horvath 2012; L{\"u} et al. 2017).
$L_{\rm acc}=GM_{\rm NS}\dot M_{\rm NS}/R_{\rm NS}$ is the accretion luminosity released by NSs,
in which $G$, $M_{\rm NS}$, $\dot M_{\rm NS}$ and $R_{\rm NS}$ are the
gravitational constant, NS mass, the mass-accretion rate and the NS radius, respectively.
In the present work, we set $\alpha_{\rm irr}$ to be $0, 0.01$ and $0.03$,
and $R_{\rm NS}$ is assumed to be $10$\,km.
Additionally,
we assume that $F_{\rm irr}$ penetrates the companion surface through an exponential function,
i.e. $F_{\rm irr}e^{-\tau}$,
in which $\tau$ is the optical depth.

Fig.\,5 represents the comparison of the evolutionary tracks of NS+He star systems
with different irradiation efficiencies,
in which we set $f=0.01$ and the initial He star mass to be $0.32\,\rm M_\odot$.
We assume that the companion is illuminated by X-ray during the pre-evaporation stage.
When the mass-transfer rate is less than $\sim 5 \times 10^{-10}\,\rm M_\odot \rm yr^{-1}$ on the declining stage,
the X-ray irradiation feedback ends and the evaporation turns on.
From this figure,
we can see that the X-ray irradiation feedback hardly affects
the evolutionary tracks of the evaporation process in NS+He star systems.
\subsection{Formation of isolated MSPs}
BWs have been proposed as an alternative formation channel for producing isolated MSPs
(e.g. Kluzniak et al. 1988; van den Heuvel \& van Paradijs 1988).
However,
the standard LMXB channel cannot reproduce the isolated MSPs within the Hubble time
(e.g. van Haaften et al. 2012; Chen et al. 2013).
Ginzburg \& Quataert (2020) recently assumed that the ablation wind couples with the magnetic field of the companion
to remove the orbit angular momentum,
thereby maintaining a stable RLOF.
Their simulations indicate that the evaporation time-scales in many BWs are longer than the Hubble time,
and they claimed that evaporation alone cannot explain most of the isolated MSPs originating from BWs.

In our simulations, the companion mass can decrease to $\textless0.001\,\rm M_\odot$
or even $\textless 10^{-5}\,\rm M_\odot$ within the Hubble time (see Fig.\,4).
During the evaporation process,
the radius of companion star increases as the mass decreases,
because the companion reaches a degenerate state (see Fig.\,6 in Wang et al. 2021).
Thus, the mass-loss of companion caused by the pulsar radiation will not be halted,
although the binary orbit expands gradually as the companion loses its mass.
Ruderman \& Shaham (1985) suggested that the He degenerate companion will undergo the tidal disruption
before the companion mass decrease to $0.004\,\rm M_\odot$.
Therefore,
we speculate that the BWs through the UCXB channel with He star companions
are the potential progenitors of isolated MSPs.

To investigate the number of isolated MSPs through the UCXB channel,
we arbitrarily assume that $N_{\rm BW}/T_{\rm BW}=N_{\rm IMSP}/T_{\rm IMSP}$,
in which $N_{\rm BW}$ is the number of BWs in region (B),
i.e. $N_{\rm BW}=6$.
$N_{\rm IMSP}$ is the number of isolated MSPs through this channel.
$T_{\rm BW}$ is the life-times of BWs in region (B),
and we adopted $T_{\rm BW}\sim$ $1-7$\,Gyr (see Table 2).
$T_{\rm IMSP}$ is the life-times of isolated MSPs through this channel,
and we roughly estimated $T_{\rm IMSP}$ from ($13.7-7$) to ($13.7-1$)\,Gyr,
where 13.7\,Gyr is the age of the current Universe (Jarosik et al. 2011).
Thus, we can obtain $N_{\rm IMSP}$ $\sim 10-80$.
It is worth noting that $N_{\rm IMSP}$ strongly depends on the mass-loss rate,
that is, higher mass-loss rate could result in shorter life-times of BWs,
thereby more isolated MSPs are produced.
In addition, several alternative channels can also produce isolated MSPs,
such as the merging of an NS and a massive WD, the high-mass X-ray binary channel, and the triple-star channel, etc
(e.g. van den Heuvel \& Bonsema 1984; Camilo, Nice \& Taylor 1993; Freire et al. 2011).

\subsection{Are RBs the progenitors of BWs with $M_2$ $\lesssim 0.01\,\rm M_\odot$?}
Previous studies mainly discussed whether RBs could evolve into BWs with $M_2$ $\sim0.01-0.05\,\rm M_\odot$
(e.g. Chen et al. 2013; Benvenuto, De Vito \& Horvath 2014; Jia \& Li 2015).
Chen et al. (2013) suggested that RBs and BWs originate from different binary evolutionary tracks,
in which low evaporation efficiency can lead to the formation of BWs.
However,
Benvenuto, De Vito \& Horvath (2014) argued that RBs with relatively compact orbits
are the progenitors of BWs by considering the X-ray irradiation feedback of MS stars.
Jia \& Li (2015) pointed out that the compact RBs can evolve into BWs if a suitable $f$-value is selected.
In our simulations,
the BWs with $M_2$ $\lesssim 0.01\,\rm M_\odot$ originate from
the UCXB channel with He star companions.
This means that the BW progenitors appear as X-ray sources rather than radio pulsars.
Meanwhile,
the evolutionary tracks of the binaries do not pass through the RB region in the $M_2-P_{\rm orb}$ diagram (see Fig.\,4),
indicating that the BWs with $M_2$ $\lesssim 0.01\,\rm M_\odot$
in region (B)
may not be formed from the evolution of RBs.
\section{Summary}
In the present work,
we investigated the UCXB channel with He star companions to explain the formation of BWs
with $M_2$ $\lesssim 0.01\,\rm M_\odot$.
By using the stellar evolution code MESA,
we calculated the long-term evolution of NS+He star systems that undergo the UCXB phase
with different He star masses and evaporation efficiencies.
We summarize the main results as follows:
\begin{enumerate}
	\item [(1)]
We found that the UCXB channel with He star companions can explain
the origin of BWs with $M_2$ $\lesssim 0.01\,\rm M_\odot$
within the Hubble time,
especially for PSRs J1719-1438, J2322-2650 and J1311-3430.
	\item [(2)]
The mass of He star companions can decrease to $\textless 0.001\,\rm M_\odot$
or even $\textless10^{-5}\,\rm M_\odot$ within the Hubble time,
indicating that the NS+He star systems undergoing the UCXB stage
are the potential progenitors of isolated MSPs.
	\item [(3)]
The X-ray irradiation feedback hardly affects the evaporation process during the evolution of NS+He star systems.
	\item [(4)]
The simulations indicate that the BWs with $M_2$ $\lesssim 0.01\,\rm M_\odot$
have different formation channels
with another sub-type of BWs (with $M_2$ $\sim 0.01-0.05\,\rm M_\odot$).
	\item [(5)]
RBs may not be the progenitors of the BWs with $M_2$ $\lesssim 0.01\,\rm M_\odot$.
\end{enumerate}

\section*{Acknowledgements}
We acknowledge the anonymous referee for valuable comments that help to improve the paper.
We thank Prof. Wencong Chen and Dr. Hailiang Chen for useful discussions and comments.
This work is supported by
the National Key R\&D Program of China (Nos 2021YFA1600400/1/4),
the National Natural Science Foundation of China (Nos 12090040/3 and 11873085),
the Western Light Project of CAS (No. XBZG-ZDSYS-202117), the science research grants from the China Manned Space Project (Nos CMS-CSST-2021-A13/B07), and the Yunnan Fundamental Research Projects (Nos 2019FJ001 and 202001AS070029).

\section*{Data availability}
Results will be shared on reasonable request to corresponding author.




\label{lastpage}
\end{document}